\documentstyle[prd,aps,preprint,tighten,epsfig]{revtex}

\begin{document}

\draft

\title{Calculable CP-violating Phases in the Minimal Seesaw Model \\
of Leptogenesis and Neutrino Mixing}
\author{{\bf Wan-lei Guo} ~ and ~ {\bf Zhi-zhong Xing}}
\address{CCAST (World Laboratory), P.O. Box 8730, Beijing 100080, China \\
and Institute of High Energy Physics, Chinese Academy of Sciences, \\
P.O. Box 918 (4), Beijing 100039, China
\footnote{Mailing address} \\
({\it Electronic address: guowl@mail.ihep.ac.cn;
xingzz@mail.ihep.ac.cn}) }
\maketitle

\begin{abstract}
We show that all nontrivial CP-violating phases can be determined
in terms of three lepton flavor mixing angles and the ratio of
$\Delta m^2_{\rm sun}$ to $\Delta m^2_{\rm atm}$ in the minimal
seesaw model in which the Frampton-Glashow-Yanagida (FGY) ansatz is
incorporated. This important point allows us to make very specific
predictions for the cosmological baryon number asymmetry and CP
violation in neutrino oscillations. A measurement of the smallest
neutrino mixing angle will sensitively test the FGY ansatz,
in particular in the case that three light neutrinos have a normal
mass hierarchy.
\end{abstract}


\newpage

\section{Introduction}

In the minimal standard model of electroweak interactions,
the lepton number conservation is assumed and neutrinos are exactly
massless Weyl particles. However, today's Super-Kamiokande \cite{SK},
SNO \cite{SNO},
KamLAND \cite{KM} and K2K \cite{K2K} neutrino oscillation experiments
have provided us with very strong evidence that neutrinos are
actually massive and lepton flavor mixing does exist. The most
economical modification of the minimal standard model, which can
both accommodate neutrino masses and allow lepton number violation to
explain the cosmological baryon asymmetry via leptogenesis \cite{FY},
is to introduce two heavy right-handed neutrinos $N_{1,2}$ and
keep the Lagrangian of electroweak interactions invariant under the
$\rm SU(2)_L \times U(1)_Y$ gauge transformation \cite{FGY,Tanimoto}.
After the spontaneous electroweak symmetry breaking, this simple
but interesting model yields the following neutrino mass term:
\begin{equation}
- {\cal L}_{\rm mass} \; =\;
\overline{(\nu_e, \nu_\mu, \nu_\tau)} ~ M_{\rm D}
\left ( \matrix{ N_1 \cr N_2 \cr} \right ) +
\frac{1}{2} \overline{(N^{\rm c}_1, N^{\rm c}_2)} ~ M_{\rm R}
\left ( \matrix{ N_1 \cr N_2 \cr} \right ) ~ + ~ {\rm h.c.} \; ,
%       (1)
\end{equation}
where $N^{\rm c}_i \equiv C \bar{N}^T_i$ with $C$ being
the charge-conjugation operator; and $(\nu_e, \nu_\mu, \nu_\tau)$
denote the left-handed neutrinos. The Dirac neutrino mass matrix
$M_{\rm D}$ is a $3\times 2$ rectangular matrix, and the Majorana
neutrino mass matrix $M_{\rm R}$ is a $2\times 2$ symmetric matrix.
The scale of $M_{\rm D}$ is characterized by the electroweak scale
$v = 174$ GeV. In contrast, the scale of $M_{\rm R}$ can be much
higher than $v$, because $N_1$ and $N_2$ are $\rm SU(2)_L$ singlets
and their corresponding mass term is not subject to the scale of
gauge symmetry breaking. Then one may obtain the effective
(light and left-handed) neutrino mass matrix $M_\nu$ via the well-known
seesaw mechanism \cite{SS}:
\begin{equation}
M_\nu \; \approx \; M_{\rm D} M^{-1}_{\rm R} M^T_{\rm D} \; .
%       (2)
\end{equation}
Without loss of generality, both $M_{\rm R}$ and the charged
lepton mass matrix $M_l$ can be taken to be diagonal, real and
positive; i.e., $M_{\rm R} = {\rm Diag}\{M_1, M_2\}$ and
$M_l = {\rm Diag}\{m_e, m_\mu, m_\tau\}$, where $M_{1,2}$ denote
the masses of two heavy Majorana neutrinos. In such a specific flavor
basis, the low-energy neutrino phenomenology is governed by $M_\nu$,
while the cosmological baryon number asymmetry is associated with
$M_{\rm D}$ via the leptogenesis mechanism.

Unfortunately, the model itself has no restriction on the structure of
$M_{\rm D}$. In Ref. \cite{FGY}, Frampton, Glashow and Yanagida (FGY)
have conjectured that $M_{\rm D}$ may take the form
\begin{equation}
M_{\rm D} \; =\; \left ( \matrix{
a & ~ {\bf 0} \cr
a' & ~ b \cr
{\bf 0} & ~ b' \cr} \right ) \; ,
%       (3)
\end{equation}
or
\begin{equation}
M_{\rm D} \; =\; \left ( \matrix{
a & ~ {\bf 0} \cr
{\bf 0} & ~ b \cr
a' & ~ b' \cr} \right ) \; .
%       (4)
\end{equation}
The texture zeros in $M_{\rm D}$ could stem from an underlying
horizontal flavor symmetry. With the help of Eq. (2), one may
straightforwardly arrive at
\begin{equation}
M_\nu \; =\; \left ( \matrix{
\displaystyle \frac{a^2}{M_1} &
\displaystyle \frac{a a'}{M_1} &
{\bf 0} \cr\cr
\displaystyle \frac{a a'}{M_1} &
\displaystyle \frac{(a')^2}{M_1} + \frac{b^2}{M_2} &
\displaystyle \frac{b b'}{M_2} \cr\cr
{\bf 0} &
\displaystyle \frac{b b'}{M_2} &
\displaystyle \frac{(b')^2}{M_2} \cr} \right ) \;
%       (5)
\end{equation}
from Eq. (3); or
\begin{equation}
M_\nu \; =\; \left ( \matrix{
\displaystyle \frac{a^2}{M_1} &
{\bf 0} &
\displaystyle \frac{a a'}{M_1} \cr\cr
{\bf 0} &
\displaystyle \frac{b^2}{M_2} &
\displaystyle \frac{b b'}{M_2} \cr\cr
\displaystyle \frac{aa'}{M_1} &
\displaystyle \frac{b b'}{M_2} &
\displaystyle \frac{(a')^2}{M_1} + \frac{(b')^2}{M_2} \cr} \right ) \;
%       (6)
\end{equation}
from Eq. (4). Note that ${\rm Det}(M_\nu) = 0$ holds in either case
%%%%%%%%%%%%%%%%%
\footnote{It has been shown in Ref. \cite{X03} that
 ${\rm Det}(M_\nu) = 0$ is independent of the specific texture zeros
taken in Eq. (3) or (4). In other words, the minimal seesaw model
itself guarantees that ${\rm Det}(M_\nu) = 0$ holds automatically.}.
%%%%%%%%%%%%%%%%%
Note also that $|{\rm Det}(M_\nu)| = m_1 m_2 m_3$ holds in the
chosen flavor basis, where $m_i$ (for $i=1,2,3$) denote the masses
of three light neutrinos. Thus one of three neutrino masses must
vanish. As the solar neutrino oscillation data have set
$m_2 > m_1$ \cite{SNO}, we are left with either $m_1=0$
(normal hierarchy) or $m_3 =0$ (inverted hierarchy). In Ref. \cite{X03},
the Majorana neutrino mass matrix with one texture zero and one
vanishing eigenvalue has been classified and discussed in some detail.

The main purpose of this paper is to reveal a very striking feature of
the minimal seesaw model in which the FGY ansatz is incorporated: all
nontrivial CP-violating phases can be calculated in terms of the
lepton flavor mixing angles $(\theta_x, \theta_y, \theta_z)$ and the
ratio of $\Delta m^2_{\rm sun}$ to $\Delta m^2_{\rm atm}$, where
$\Delta m^2_{\rm sun}$ and $\Delta m^2_{\rm atm}$ stand respectively
for the typical mass-sqaured differences of solar and atmospheric
neutrino oscillations. This important point, which was not observed
in the previous analyses of the minimal seesaw model \cite{FGY,Tanimoto,RS},
implies that a stringent test of the FGY ansatz can simply be realized
once the smallest mixing angle $\theta_z$ is measured or constrained
to a better degree of accuracy. Considering both normal and inverted
mass hierarchies of three light neutrinos, we obtain very specific
predictions for the cosmological baryon number asymmetry, the effective
mass of neutrinoless double beta decay and CP violation in neutrino
oscillations.

\section{Determination of CP-violating phases}

Without loss of generality, one may always redefine the phases of
charged lepton fields to make $a$, $b$ and $b'$ of $M_{\rm D}$ real
and positive \cite{FGY}. In other words, only $a'$ is complex and
its phase $\phi \equiv \arg(a')$ is the sole source of CP violation
in the model under discussion. Because both $M_l$ and $M_{\rm R}$
have been taken to be diagonal, real and positive, $M_\nu$ can in
general be parametrized as follows:
\begin{equation}
M_\nu \; =\; \left (PUQ \right ) \left ( \matrix{
m_1 & 0 & 0 \cr
0 & m_2 & 0 \cr
0 & 0 & m_3 \cr} \right ) \left (PUQ \right )^T \; ,
%       (7)
\end{equation}
where $P= {\rm Diag}\{e^{i\alpha}, e^{i\beta}, e^{i\gamma}\}$ and
$Q= {\rm Diag}\{e^{i\rho}, e^{i\sigma}, e^{i\omega}\}$ are two
phase matrices, and $U$ is given by
\begin{equation}
U \; = \; \left ( \matrix{
c_x c_z & s_x c_z & s_z \cr
- c_x s_y s_z - s_x c_y e^{-i\delta} &
- s_x s_y s_z + c_x c_y e^{-i\delta} &
s_y c_z \cr
- c_x c_y s_z + s_x s_y e^{-i\delta} &
- s_x c_y s_z - c_x s_y e^{-i\delta} &
c_y c_z \cr } \right ) \;
%       (8)
\end{equation}
with $s_x \equiv \sin\theta_x$, $c_x \equiv \cos\theta_x$ and so
on. The phase parameter of $U$ (Dirac phase) governs the strength of
CP violation in neutrino oscillations, while two independent phase
parameters of $Q$ (Majorana phases) are relevant to the neutrinoless
double beta decay \cite{FX01}. The phases
of $P$ cannot be neglected in the parametrization of $M_\nu$ ---
their essential role is to fulfil a complete match between the phases
of $M_\nu$ in Eq. (5) or (6) and those defined in Eqs. (7) and (8).
It is then obvious that the nontrivial phases of $P$, $U$ and $Q$
should have definite relations with $\phi$.

Note that three mixing angles of $U$ can directly be given
in terms of the mixing angles of solar, atmospheric and
reactor \cite{CHOOZ} neutrino oscillations. Namely,
$\theta_x \approx \theta_{\rm sun}$, $\theta_y \approx \theta_{\rm atm}$
and $\theta_z \approx \theta_{\rm chz}$ hold as a good approximation.
In view of current experimental data, we have
$\theta_x \approx 32^\circ$ and $\theta_y \approx 45^\circ$ (best-fit
values \cite{FIT}) as well as $\theta_z < 12^\circ$. The mass-squared
differences of solar and atmospheric neutrino oscillations are defined
respectively as $\Delta m^2_{\rm sun} \equiv m^2_2 - m^2_1$ and
$\Delta m^2_{\rm atm} \equiv |m^2_3 - m^2_2|$. Their best-fit values
read $\Delta m^2_{\rm sun} \approx 7.13 \times 10^{-5} ~ {\rm eV}^2$ and
$\Delta m^2_{\rm atm} \approx 2.6 \times 10^{-3} ~ {\rm eV}^2$ \cite{FIT}.
These typical numbers will be used in our numerical calculations.

To be more specific, we shall concentrate on the FGY ansatz for $M_{\rm D}$
in Eq. (3) or equivalently $M_\nu$ in Eq. (5). Some brief comments will
be given on the consequences of $M_{\rm D}$ in Eq. (4) or equivalently
$M_\nu$ in Eq. (6). Indeed, both possibilities lead to very similar
phenomenological results.

\subsection{Normal neutrino mass hierarchy ($m_1 =0$)}

If $m_1 =0$ holds, we obtain
$m_2 =\sqrt{\Delta m^2_{\rm sun}} \approx 8.4 \times 10^{-3}$ eV and
$m_3 =\sqrt{\Delta m^2_{\rm sun} + \Delta m^2_{\rm atm}}
\approx 5.2 \times 10^{-2}$ eV. In this case, only a single Majorana
phase of CP violation is physically nontrivial. Hence the phase matrix
$Q$ can be simplied to $Q = {\rm Diag}\{ 1, e^{i\sigma}, 1\}$ and
six independent matrix elements of $M_\nu$ can be written as
\begin{eqnarray}
(M_\nu)_{11} & = & e^{2i\alpha} \left [m_2 s^2_x c^2_z e^{2i\sigma}
+ m_3 s^2_z \right ] \; ,
\nonumber  \\
(M_\nu)_{22} & = & e^{2i\beta} \left [m_2 \left (-s_x s_y s_z + c_x c_y
e^{-i\delta} \right )^2 e^{2i\sigma} + m_3 s^2_y c^2_z \right ] \; ,
\nonumber \\
(M_\nu)_{33} & = & e^{2i\gamma} \left [m_2 \left (s_x c_y s_z +
c_x s_y e^{-i\delta} \right )^2 e^{2i\sigma} + m_3 c^2_y c^2_z \right ] \; ;
%       (9)
\end{eqnarray}
and
\begin{eqnarray}
(M_\nu)_{12} & = & e^{i(\alpha + \beta)} \left [ m_2 s_x c_z
\left (-s_x s_y s_z + c_x c_y e^{-i\delta} \right ) e^{2i\sigma} +
m_3 s_y s_z c_z \right ] \; ,
\nonumber  \\
(M_\nu)_{13} & = & e^{i(\alpha+\gamma)} \left [ -m_2 s_x c_z
\left (s_x c_y s_z + c_x s_y e^{-i\delta} \right ) e^{2i\sigma}
+ m_3 c_y s_z c_z \right ] \; ,
\nonumber \\
(M_\nu)_{23} & = & e^{i(\beta+\gamma)} \left [ -m_2
\left (s_x c_y s_z + c_x s_y e^{-i\delta} \right )
\left (-s_x s_y s_z + c_x c_y e^{-i\delta} \right ) e^{2i\sigma}
+ m_3 s_y c_y c^2_z \right ] \; .
%       (10)
\end{eqnarray}
Because of $(M_\nu)_{13}=0$ as shown in Eq. (5), we straightforwardly
obtain
\begin{eqnarray}
\delta & = & \pm \arccos \left [ \frac{c^2_y s^2_z - \xi^2 s^2_x
\left (c^2_x s^2_y + s^2_x c^2_y s^2_z \right )}{2 \xi^2 s^3_x c_x
s_y c_y s_z} \right ] \; ,
\nonumber \\
\sigma & = & \frac{1}{2} \arctan \left [\frac{c_x s_y \sin\delta}
{s_x c_y s_z + c_x s_y \cos\delta} \right ] \; ,
%       (11)
\end{eqnarray}
where $\xi \equiv m_2/m_3 \approx 0.16$. This result implies that both
$\delta$ and $\sigma$ can definitely be determined, if and only if the
smallest mixing angle $\theta_z$ is measured.

To establish the relationship between $\phi$ and $\delta$, we need to
figure out $\alpha$, $\beta$ and $\gamma$. Because $a$, $b$ and $b'$
are real and positive, $(M_\nu)_{11}$, $(M_\nu)_{23}$ and $(M_\nu)_{33}$
must be real and positive. Then $\alpha$, $\beta$ and $\gamma$ can
be derived from Eqs. (10) and (11):
\begin{eqnarray}
\alpha & = & -\frac{1}{2} \arctan \left [ \frac{\xi s^2_x c^2_z
\sin 2\sigma}{s^2_z + \xi s^2_x c^2_z \cos 2\sigma} \right ] \; ,
\nonumber \\
\beta & = & -\gamma - \arctan \left [ \frac{c_x c_y s_z \sin\delta}
{s_x s_y - c_x c_y s_z \cos\delta} \right ] \; ,
\nonumber \\
\gamma & = & +\frac{1}{2} \arctan \left [ \frac{s^2_z
\sin 2\sigma}{\xi s^2_x c^2_z + s^2_z \cos 2\sigma} \right ] \; .
%       (12)
\end{eqnarray}
Then the overall phase of $(M_\nu)_{12}$, which is equal to the
phase of $a'$, is given by
\begin{equation}
~~~ \phi \; =\; \alpha + \beta - \arctan \left [ \frac{s_x c_y s_z
\sin\delta}{c_x s_y + s_x c_y s_z \cos\delta} \right ] \; .
%       (13)
\end{equation}
Now that all six phase parameters ($\delta$, $\sigma$,
$\phi$, $\alpha$, $\beta$ and $\gamma$) have been determined in terms of
$\xi$, $\theta_x$, $\theta_y$ and $\theta_z$, a measurement of the
unknown angle $\theta_z$ becomes crucial to test the model.

\subsection{Inverted neutrino mass hierarchy ($m_3 =0$)}

If $m_3 =0$ holds, we arrive at
$m_1 =\sqrt{\Delta m^2_{\rm atm} - \Delta m^2_{\rm sun}}
\approx 5.0 \times 10^{-2}$ eV and $m_2 =\sqrt{\Delta m^2_{\rm atm}}
\approx 5.1 \times 10^{-2}$ eV. Since only a single Majorana phase
of CP violation is physically nontrivial, we can always simplify the
phase matrix $Q$ to $Q={\rm Diag}\{1, e^{i\sigma}, 1\}$. In this case,
six independent matrix elements of $M_\nu$ turn out to be
\begin{eqnarray}
(M_\nu)_{11} & = & e^{2i\alpha} \left [m_1 c^2_x c^2_z
+ m_2 s^2_x c^2_z e^{2i\sigma} \right ] \; ,
\nonumber  \\
(M_\nu)_{22} & = & e^{2i\beta} \left [m_1 \left (c_x s_y s_z + s_x c_y
e^{-i\delta} \right )^2  + m_2 \left (-s_x s_y s_z + c_x c_y e^{-i\delta}
\right )^2 e^{2i\sigma} \right ] \; ,
\nonumber \\
(M_\nu)_{33} & = & e^{2i\gamma} \left [m_1 \left (-c_x c_y s_z +
s_x s_y e^{-i\delta} \right )^2 + m_2 \left (s_x c_y s_z + c_x s_y
e^{-i\delta} \right )^2 e^{2i\sigma} \right ] \; ;
%       (14)
\end{eqnarray}
and
\begin{eqnarray}
(M_\nu)_{12} & = & e^{i(\alpha + \beta)} \left [ -m_1 c_x c_z
\left (c_x s_y s_z + s_x c_y e^{-i\delta} \right ) +
m_2 s_x c_z \left (-s_x s_y s_z + c_x c_y e^{-i\delta} \right )
e^{2i\sigma} \right ] \; ,
\nonumber  \\
(M_\nu)_{13} & = & e^{i(\alpha+\gamma)} \left [ m_1 c_x c_z
\left (-c_x c_y s_z + s_x s_y e^{-i\delta} \right ) - m_2 s_x c_z
\left (s_x c_y s_z + c_x s_y e^{-i\delta} \right )
e^{2i\sigma} \right ] \; ,
\nonumber \\
(M_\nu)_{23} & = & e^{i(\beta+\gamma)} \left [ -m_1
\left (c_x s_y s_z + s_x c_y e^{-i\delta} \right )
\left (-c_x c_y s_z + s_x s_y e^{-i\delta} \right ) \right .
\nonumber \\
&  & \left . ~~~~~~~~~ - m_2
\left (-s_x s_y s_z + c_x c_y e^{-i\delta} \right )
\left (s_x c_y s_z + c_x s_y e^{-i\delta} \right )
e^{2i\sigma} \right ] \; .
%       (15)
\end{eqnarray}
Because of $(M_\nu)_{13}=0$ as shown in Eq. (5), we get
\begin{eqnarray}
\delta & = & \pm \arccos \left [ \frac{\left (\zeta^2 c^4_x -
s^4_x \right ) c^2_y s^2_z + \left (\zeta^2 - 1 \right ) s^2_x
c^2_x s^2_y}{2 s_x c_x \left (s^2_x + \zeta^2 c^2_x \right )
s_y c_y s_z} \right ] \; ,
\nonumber \\
\sigma & = & -\frac{1}{2} \arctan \left [\frac{s_y c_y s_z \sin\delta}
{s_x c_x \left (s^2_y - c^2_y s^2_z \right ) +
\left (s^2_x - c^2_x \right ) s_y c_y s_z \cos\delta} \right ] \; ,
%       (16)
\end{eqnarray}
where $\zeta \equiv m_1/m_2 \approx 0.98$. Once the smallest mixing
angle $\theta_z$ is observed, one may determine both $\delta$ and
$\sigma$ with the help of Eq. (16).

The phase parameters $\alpha$, $\beta$ and $\gamma$ can be fixed
by taking account of the positiveness of $(M_\nu)_{11}$, $(M_\nu)_{23}$
and $(M_\nu)_{33}$. With the help of Eqs. (14) and (15), we obtain
\begin{eqnarray}
\alpha & = & -\frac{1}{2} \arctan \left [ \frac{s^2_x \sin 2\sigma}
{\zeta c^2_x + s^2_x \cos 2\sigma} \right ] \; ,
\nonumber \\
\beta & = & \gamma + \pi \; ,
\nonumber \\
\gamma & = & \frac{\delta}{2} + \frac{1}{2} \arctan \left [
\frac{s_x s_y \sin \delta}{s_x s_y \cos \delta - c_x c_y s_z} \right ]
- \pi \; .
%       (17)
\end{eqnarray}
Then the overall phase of $(M_\nu)_{12}$, which is equal to the
phase of $a'$, is given by
\begin{equation}
\phi \; =\; \alpha + \beta -
\arctan \left [ \frac{s_x c_y s_z \sin\delta}{c_x s_y +
s_x c_y s_z \cos\delta} \right ] - \pi \; .
%       (18)
\end{equation}
Again, a measurement of the unknown mixing angle $\theta_z$ will allow
us to determine all six phase parameters ($\delta$, $\sigma$,
$\phi$, $\alpha$, $\beta$ and $\gamma$).

\subsection{Numerical dependence of $(\delta, \sigma, \phi,
\alpha, \beta, \gamma)$ on $\theta_z$}

Using the best-fit values of $\Delta m^2_{\rm sun}$,
$\Delta m^2_{\rm atm}$, $\theta_x$ and $\theta_y$, we illustrate the
numerical dependence of six phase parameters
$(\delta, \sigma, \phi, \alpha, \beta, \gamma)$ on the smallest
mixing angle $\theta_z$ in Fig. 1(a) and Fig. 1(b) for the
$m_1 =0$ case and in Fig. 2(a) and Fig. 2(b) for the $m_3 =0$ case.
Some discussions are in order.

(1) In the $m_1=0$ case, $\theta_z$ is restricted to a very narrow
range $4.0^\circ \lesssim \theta_z \lesssim 4.4^\circ$ (namely,
$0.070 \lesssim s_z \lesssim 0.077$). This result implies that the
FGY ansatz with $m_1=0$ is highly sensitive to $\theta_z$ and can
easily be ruled out if the experimental value of $\theta_z$ does
not really lie in the predicted region. To a good degree of
accuracy, we obtain $\delta \approx 2\sigma$, $\phi \approx \alpha
\approx -\sigma$, $\beta \approx -\gamma$, and $\gamma \approx 0$.
These instructive relations can essentially be observed from Eqs.
(11), (12) and (13), because of $s_z \ll 1$. Note that we have
only shown the dependence of $\delta$ on $\theta_z$ in the range
$0< \delta <\pi$. The reason is simply that only this range can
lead to $Y_{\rm B} >0$ (i.e., the positive cosmological baryon
number asymmetry), as one can see later on.

(2) In the $m_3=0$ case, there is no strong constraint on $\theta_z$
except that $\theta_z > 0.36^\circ$ (or equivalently $s_z > 0.0063$)
must hold. We see that $\delta \approx \beta \approx \phi + \pi$,
$\phi \approx \gamma$, $\sigma \approx -\alpha$ and $\alpha \approx 0$
hold to a good degree of accuracy. These results can also be
observed from Eqs. (16), (17) and (18) by taking account of $s_z \ll 1$.
Again, we have used the positive sign of $Y_{\rm B}$ to constrain the
allowed range of $\delta$. The other phase parameters are required to
take possible values between $-\pi$ and $+\pi$.

As useful by-products, the Jarlskog parameter of CP violation
($J_{\rm CP}$ \cite{J}) and the effective mass of neutrinoless
double beta decay ($\langle m\rangle_{ee}$ \cite{X03R}) can be
calculated. We show the numerical results of $J_{\rm CP}$ versus
$\langle m\rangle_{ee}$ in Fig. 1(c) for $m_1=0$ and in Fig. 2(c)
for $m_3 =0$, respectively. It is clear that $0 < J_{\rm CP}
\lesssim 0.016$ and $2.1 ~ {\rm meV} \lesssim \langle m
\rangle_{ee} \lesssim 2.7$ meV in the $m_1=0$ case, while $0 <
J_{\rm CP} \lesssim 0.035$ and $\langle m \rangle_{ee} \approx
\sqrt{\Delta m^2_{\rm atm}} \approx 0.05$ eV in the $m_3=0$ case.
The present experimental upper bound of $\langle m \rangle_{ee}$
is $\langle m \rangle_{ee} < 0.35$ eV at the $90\%$ confidence
level \cite{HM}.

\section{Baryon asymmetry via leptogenesis}

Because of lepton number violation, two heavy Majorana neutrinos
$N_i$ (for $i=1$ and 2) may decay into $lH$ and its CP-conjugate
state, where $l$ denotes the left-handed lepton doublet and $H$
stands for the Higgs-boson weak isodoublet. The decay occurs at both
the tree level and the one-loop level (via self-energy and vertex
corrections), and their interference leads to a CP-violating
asymmetry $\varepsilon_i$ between the $CP$-conjugated
$N_i \rightarrow l + H$ and $N_i \rightarrow \bar{l} + H^*$
processes \cite{FY}. If the masses of $N_1$ and $N_2$ are hierarchical
(i.e., $M_1 \ll M_2$), the interactions of $N_1$ can be in thermal
equilibrium when $N_2$ decays. The asymmetry $\varepsilon_2$
is therefore erased before $N_1$ decays, and only the asymmetry
$\varepsilon_1$ produced by the out-of-equilibrium decay
of $N_1$ survives. In the flavor basis chosen above, we have
\begin{eqnarray}
\varepsilon_1 & \equiv & \frac{\Gamma (N_1 \rightarrow l + H)
~ - ~ \Gamma (N_1 \rightarrow \bar{l} + H^*)}
 {\Gamma (N_1 \rightarrow l + H)
~ + ~ \Gamma (N_1 \rightarrow \bar{l} + H^*)}
\nonumber \\
& \approx & -\frac{3}{16\pi v^2} \cdot \frac{M_1}{M_2} \cdot
\frac{{\rm Im} \left [ (M^\dagger_{\rm D} M_{\rm D})_{12} \right ]^2}
{(M^\dagger_{\rm D} M_{\rm D})_{11}}
\nonumber \\
& = & \frac{3}{16\pi v^2} \cdot
\frac{\displaystyle M_1 |(M_\nu)_{12}|^2 |(M_\nu)_{23}|^2 \sin 2\phi}
{\displaystyle \left [ |(M_\nu)_{11}|^2 + |(M_\nu)_{12}|^2 \right ]
|(M_\nu)_{33}|} \;\; .
%       (19)
\end{eqnarray}
In deriving this formula, we have used
$(M^\dagger_{\rm D} M_{\rm D})_{11} = a^2 + |a'|^2$ and
$(M^\dagger_{\rm D} M_{\rm D})_{12} = (a')^*b$ as well as
\begin{eqnarray}
a^2 & = & M_1 |(M_\nu)_{11}| \; , ~~~~
|a'|^2 =  M_1 \frac{|(M_\nu)_{12}|^2}{|(M_\nu)_{11}|} \; ,
\nonumber \\
b^2 & = & M_2 \frac{|(M_\nu)_{23}|^2}{|(M_\nu)_{33}|} \; , ~~~
(b')^2 = M_2 |(M_\nu)_{33}| \;\; .
%       (20)
\end{eqnarray}
One can see that $\varepsilon_1$ is independent of $M_2$, as long as
$M_2 \gg M_1$ is satisfied. A nonvanishing $\varepsilon_1$ may result in
a net lepton number asymmetry $Y_{\rm L} \equiv n^{~}_{\rm L}/{\bf s} =
d\varepsilon_1/g^{~}_*$, where $g^{~}_* = 106.75$
is an effective number characterizing the relativistic degrees of freedom
which contribute to the entropy {\bf s} of the early universe, and $d$
accounts for the dilution effects induced by the lepton-number-violating
wash-out processes \cite{R}. If the effective neutrino mass parameter
$\tilde{m}_1 \equiv (M^\dagger_{\rm D} M_{\rm D})_{11}/M_1$ \cite{BP}
lies in the range
$10^{-2} ~ {\rm eV} \lesssim \tilde{m}_1 \lesssim 10^3 ~ {\rm eV}$, one 
may estimate the value of $d$ by use of the following approximate
formula \cite{Kolb}
%%%%%%%%%%%%%%%%%%%%%%%
\footnote{For $M_1 \ll 10^{14}$ GeV, Giudice {\it et al} have presented
a different approximate formula for the dilution factor (denoted as
$\eta$ \cite{Raidal}):
$\displaystyle \frac{1}{\eta} \approx \frac{3.3 \times 10^{-3} ~ {\rm eV}}
{\tilde{m_1}} + \left (\frac{\tilde{m}_1}{5.5 \times 10^{-4} ~ {\rm eV}}
\right )^{1.16}$. We find that the ratio $\eta/d$ will vary in the range
$0.9 \lesssim \eta/d \lesssim 2.0$, if $\tilde{m}_1$ takes values in the 
region $10^{-2} ~ {\rm eV} \lesssim \tilde{m}_1 \lesssim 10^3 ~ {\rm eV}$.
Thus there is no significant inconsistency between two empirical formulas.}:
%%%%%%%%%%%%%%%%%%%%%%
\begin{equation}
d \; \approx \; 0.3 \left (\frac{10^{-3} ~ {\rm eV}}{\tilde{m}_1} \right )
\left [ \ln \left ( \frac{\tilde{m}_1}{10^{-3} ~ {\rm eV}} \right )
\right ]^{-0.6} \; . ~~~
%       (21)
\end{equation}
The lepton number asymmetry $Y_{\rm L}$ is eventually converted
into a net baryon number asymmetry $Y_{\rm B}$ via the
nonperturbative sphaleron processes \cite{Kuzmin}: $Y_{\rm B}
\equiv n^{~}_{\rm B}/{\bf s} \approx -0.55 Y_{\rm L}$. A generous
range $0.7 \times 10^{-10} \lesssim Y_{\rm B} \lesssim 1.0 \times
10^{-10}$ has been drawn from the recent WMAP observational data
\cite{WMAP}.

It is clear that $\varepsilon_1$ and $Y_{\rm B}$ only involve two free
parameters: $M_1$ and $\phi$. Because $\phi$ is associated
with the unknown flavor mixing angle $\theta_z$, one may analyze the
dependence of $Y_{\rm B}$ on $\theta_z$ for given values of $M_1$.
For $m_1=0$ and $m_3=0$ cases, we plot the numerical results of $Y_{\rm B}$
in Fig. 1(d) and Fig. 2(d) respectively. Some comments are in order.

(1) In the $m_1=0$ case, current observational data of $Y_{\rm B}$
require $M_1 \geq 2.9 \times 10^{10}$ GeV for the allowed ranges
of $s_z$. Once $s_z$ is precisely measured, it is possible to fix
the value of $M_1$ in most cases (e.g., $M_1 = 10^{11}$ GeV will
be ruled out, if $s_z \approx 0.074$ holds).

(2) In the $m_3 =0$ case, $M_1 \geq 2.7 \times 10^{13}$ GeV is
required by current observational data of $Y_{\rm B}$. Although
$\theta_z$ is less restricted in this scenario, it remains
possible to determine the value of $M_1$ once $\theta_z$ is
measured (e.g., $M_1 \approx 5 \times 10^{13}$ GeV is expected, if
$s_z \approx 0.1$ holds).

(3) One can carry out a similar analysis of $Y_{\rm B}$ in the framework
of supersymmetric seesaw and leptogenesis models. However, the FGY
ansatz does not favor $M_1 \lesssim 10^8$ GeV, which crucially affects
the maximum reheating temperature of the universe after inflation in
the generic supergravity models \cite{R}.

We remark that the cosmological baryon number asymmetry is closely 
correlated with the Jarlskog parameter of CP violation in the FGY
ansatz
%%%%%%%%%%%%%%%%%%%
\footnote{A similar point has been discussed by Endoh 
{\it et al} \cite{Tanimoto} in a more general way for the minimal
seesaw model with leptogenesis. See, also, Ref. \cite{Branco} for
a possible seesaw bridge between leptogenesis and CP violation at
low energies beyond the minimal seesaw model.}.
%%%%%%%%%%%%%%%%%%%
For illustration, we plot the numerical correlation between
$Y_{\rm B}$ and $J_{\rm CP}$ in Fig. 3, where 
$M_1 = 5 \times 10^{10}$ GeV for the $m_1 =0$ case and
$M_1 = 5 \times 10^{13}$ GeV for the $m_3 =0$ case have typically
been taken. One can see that the observationally-allowed range
of $Y_{\rm B}$ corresponds to $J_{\rm CP} \sim 1\%$ in the
$m_1 =0$ case and $J_{\rm CP} \sim 2\%$ in the
$m_3 =0$ case. The correlation between $Y_{\rm B}$ and $J_{\rm CP}$
is so strong that a measurement of the latter in the long-baseline
neutrino oscillation experiments could shed some light on the
ball-park magnitude of $M_1$. 

It is worth mentioning that we have neglected possible
renormalization-group running effects of neutrino masses and lepton
flavor mixing parameters between the scales of $v$ and $M_1$ \cite{RGE}.
Such an approximation is expected to be safe in the $m_1=0$ case, in
which three light neutrinos have a clear mass hierarchy. In the
$m_3=0$ case, in which $m_1 \approx m_2$ holds, a careful analysis of
possible running effects on the FGY ansatz is needed \cite{GMX} but it
is beyond the scope of this paper.

Finally let us comment on possible phenomenological consequences of
the FGY ansatz for $M_{\rm D}$ in Eq. (4) or equivalently $M_\nu$ in
Eq. (6). In the $m_1 =0$ case, one can make use of Eqs. (9) and (10) to
calculate relevant phase parameters by setting $(M_\nu)_{12} =0$.
A similar analysis can be done for the $m_3 =0$ case by using
Eqs. (14) and (15) and taking $(M_\nu)_{12} =0$. We find that
the simple replacements $\delta \rightarrow \delta - \pi$ and
$\theta_y \rightarrow \pi/2 - \theta_y$ may allow us to write out the
expressions of $\sigma$, $\phi$, $\alpha$, $\beta$ and $\gamma$
in the $(M_\nu)_{12} =0$ case directly from Eqs. (11)--(13) and (16)--(18).
It turns out that the numerical results of $\sigma$, $\phi$ and $\alpha$
are essentially unchanged, but those of $\beta$, $\gamma$ and $J_{\rm CP}$
require the replacements $\beta\rightleftharpoons\gamma$ and
$J_{\rm CP} \rightarrow -J_{\rm CP}$. The results for $Y_{\rm B}$ are
essentially identical in $(M_\nu)_{12} =0$ and $(M_\nu)_{13} =0$ cases.

\section{Summary}

We have analyzed the minimal seesaw model for leptogenesis and
neutrino mixing, in which the FGY ansatz is incorporated. We point
out a very striking feature of this model: all
nontrivial CP-violating phases can be determined in terms of the
lepton flavor mixing angles and the ratio of $\Delta m^2_{\rm sun}$
to $\Delta m^2_{\rm atm}$. This important observation allows us to
make very specific and testable predictions for the cosmological
baryon number asymmetry, the effective mass of neutrinoless double
beta decay and CP violation in neutrino oscillations. A precise
measurement of the smallest mixing angle in reactor-
and accelerator-based neutrino oscillation experiments will be
extremely helpful to examine the FGY ansatz and other presently
viable ans$\rm\ddot{a}$tze of lepton mass matrices.

\acknowledgments{We are indebted to J.W. Mei for helping us to remove
an error in our original numerical calculations. We are also grateful 
to M. Raidal for very useful communication. This work was supported in 
part by the National Nature Science Foundation of China.}

\newpage

%%%%%%%%%%%%%%%%%%%% Fig. 1 %%%%%%%%%%%%%%%%
\begin{figure}[t]
\hspace{-1cm}
\epsfig{file=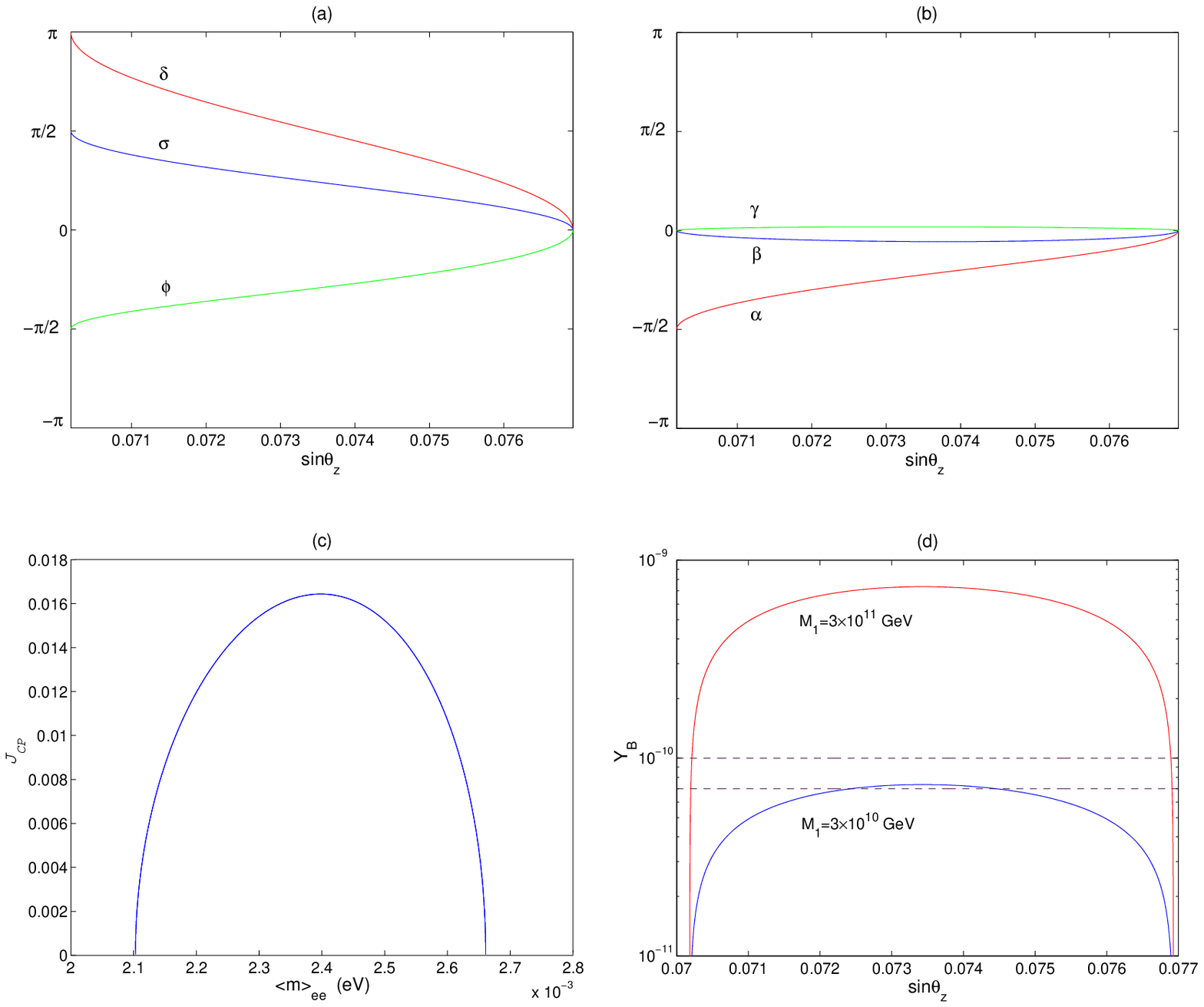,bbllx=-2cm,bblly=2cm,bburx=19cm,bbury=28cm,%
width=14.5cm,height=21cm,angle=0,clip=0} \vspace{-5cm}
\caption{Numerical results for the $m_1=0$ case: (a) dependence of
$\delta$, $\sigma$ and $\phi$ on $\sin\theta_z$; (b) dependence of
$\alpha$, $\beta$ and $\gamma$ on $\sin\theta_z$; (c) allowed ranges
of $J_{\rm CP}$ and $\langle m\rangle_{ee}$; (d) dependence of
$Y_{\rm B}$ on $M_1$ and $\sin\theta_z$. The region between two
dashed lines in (d) corresponds to the range of $Y_{\rm B}$ allowed
by current observational data.}
\end{figure}
%%%%%%%%%%%%%%%%%%%%%%%%%%%%%%%%%%%%%%%%%%%

\newpage

%%%%%%%%%%%%%%%%%%%% Fig. 2 %%%%%%%%%%%%%%%%
\begin{figure}[t]
\hspace{-1cm}
\epsfig{file=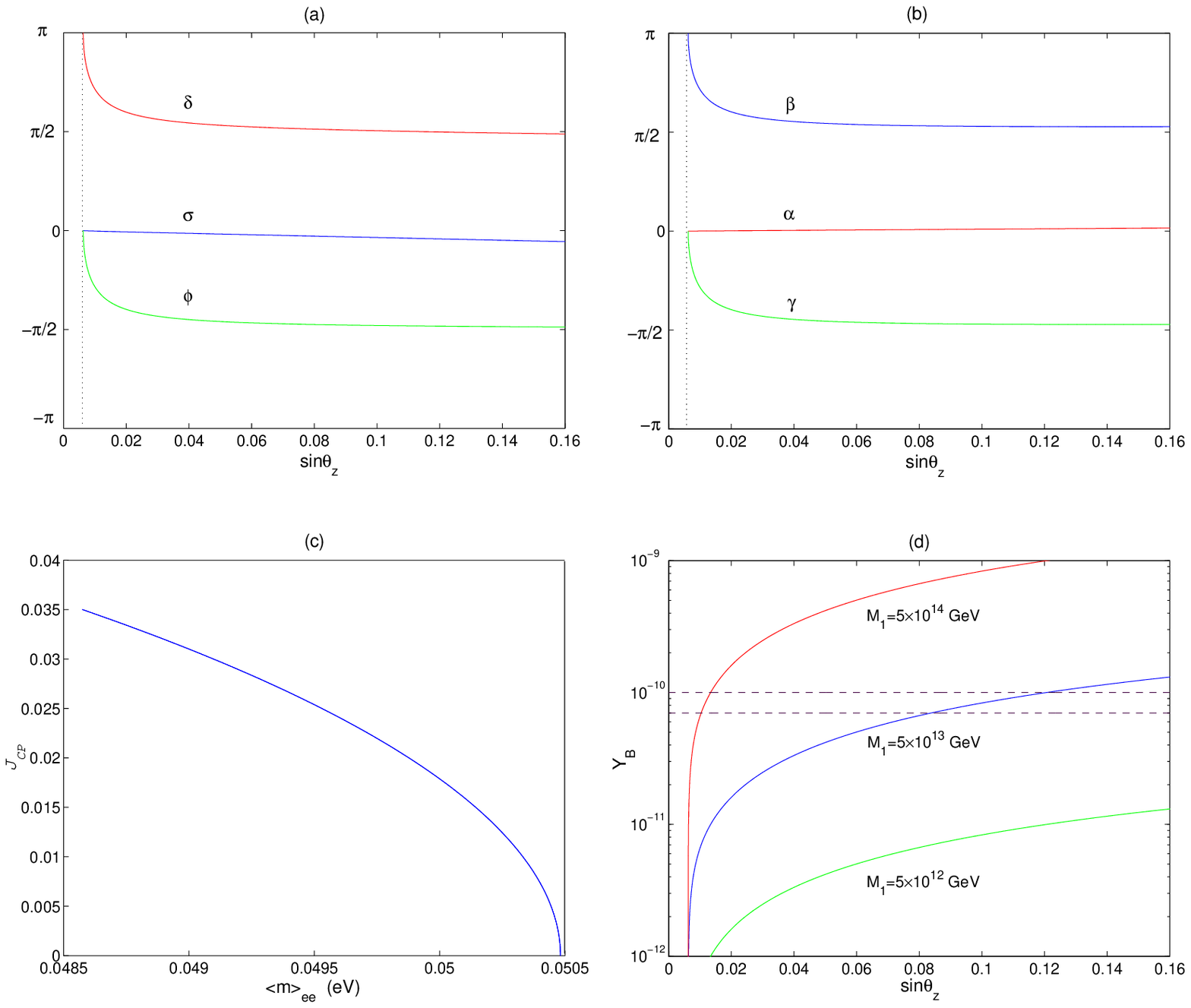,bbllx=-2cm,bblly=2cm,bburx=19cm,bbury=28cm,%
width=14.5cm,height=21cm,angle=0,clip=0} \vspace{-5cm}
\caption{Numerical results for the $m_3=0$ case: (a) dependence of
$\delta$, $\sigma$ and $\phi$ on $\sin\theta_z$; (b) dependence of
$\alpha$, $\beta$ and $\gamma$ on $\sin\theta_z$; (c) allowed ranges
of $J_{\rm CP}$ and $\langle m\rangle_{ee}$; (d) dependence of
$Y_{\rm B}$ on $M_1$ and $\sin\theta_z$. The region between two
dashed lines in (d) corresponds to the range of $Y_{\rm B}$ allowed
by current observational data.}
\end{figure}
%%%%%%%%%%%%%%%%%%%%%%%%%%%%%%%%%%%%%%%%%%%

\newpage

%%%%%%%%%%%%%%%%%%%% Fig. 3 %%%%%%%%%%%%%%%%
\begin{figure}[t]
\hspace{-1cm}
\epsfig{file=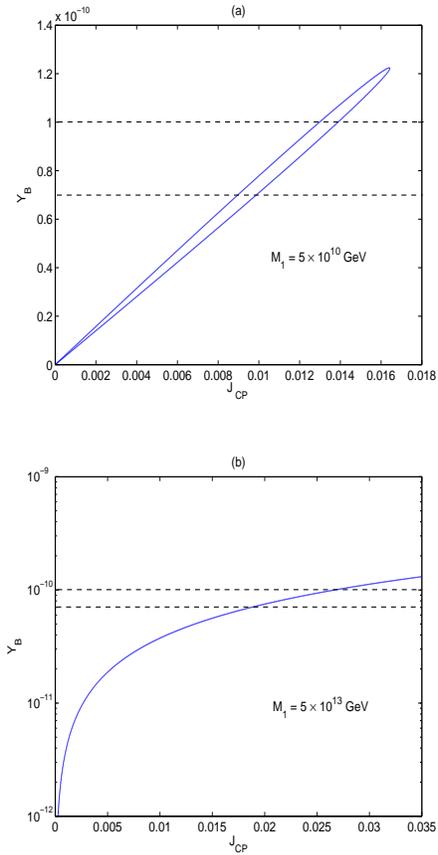,bbllx=-6.3cm,bblly=2cm,bburx=14.8cm,bbury=28cm,%
width=14.5cm,height=21cm,angle=0,clip=0} \vspace{-5cm}
\caption{Numerical illustration of the correlation between
$Y_{\rm B}$ and $J_{\rm CP}$: (a) in the $m_1 =0$ case with
$M_1 = 5 \times 10^{10}$ GeV; and
(b) in the $m_3 =0$ case with $M_1 = 5 \times 10^{13}$ GeV.
The region between two dashed lines in (a) or (b) corresponds to 
the range of $Y_{\rm B}$ allowed by current observational data.}
\end{figure}
%%%%%%%%%%%%%%%%%%%%%%%%%%%%%%%%%%%%%%%%%%%

\end{document}